\documentstyle[aps,prl,preprint,epsfig]{revtex}

\begin{document}
\preprint{TU-575}
\title{Fate of Gravitons in Warped Extra Dimension}
\author{Sanghyeon Chang and Masahiro Yamaguchi}
\address{Department of Physics, Tohoku University,
Sendai 980-8578, Japan}
\maketitle
\begin{abstract}
Interaction of bulk gravitons to Standard Model particles is examined in the 
scenario of warped extra dimension.  It is found that the length of the
extra dimension is of physical importance as it determines the decay widths 
of graviton Kaluza-Klein modes. We show that cosmology as well as collider signals of the Kaluza-Klein modes are very different for different choices of
the length of the fifth dimension.
\end{abstract} 

The higher-dimensional unified theory has been a fascinating idea
since it was first advocated by Kaluza and Klein \cite{Kaluza}. Most
of these attempts including weakly interacting string theories 
have extra dimensions as small as the Planck length, and also
predict very heavy Kaluza-Klein modes with masses close to the Planck
mass.  If this were a general consequence of the Kaluza-Klein theory,
it could not be directly accessible at collider experiments.

Recently there have been, however, 
some proposals \cite{ADD1,AADD,RS1,GO,RS2,others1,LR,others2}
to testable models of the Kaluza-Klein idea, in which the large
hierarchy between the Planck scale and the electroweak scale is
explained by the geometry of the extra dimension(s).  In
Refs. \cite{ADD1,AADD}, the large extra dimensions are utilized to
suppress the gravitational interaction while the fundamental scale is
close to the electroweak scale. The Kaluza-Klein modes in this
scenario are very light and can be probed at collider experiments etc.
This opens up a possibility of experimentally testable models of 
the Kaluza-Klein theory.

More recently it has been pointed out that an appropriate
choice of the bulk cosmological constant as well as vacuum energies at
3-branes allows us to have a non-factorizable metric along the fifth
dimension \cite{RS1,GO,RS2,others1,LR,others2}. 
This warped extra dimension gives exponential suppression
of the metric at a point far away from the Planck brane where gravitons are
localized, yielding an
intriguing interpretation of the electroweak scale. In this case
\cite{RS1,GO,RS2,LR}, the metric in 4+1 space time
becomes
\begin{equation}
ds^2= e^{-2k r_c \phi} \eta_{\mu\nu} dx^\mu dx^\nu + r_c^2 d\phi^2
\end{equation}
where $0\leq \phi\leq \pi$ and $r_c$ is the size of compactified
fifth dimension. We assume that the Planck brane is located at $\phi=0$. 
The Planck scale can be written with fundamental scales
$M, k$ and the size of the fifth dimension $r_c$.
\begin{equation}
M_{pl}^2=\frac{M^3}{k}(1 - e^{-2kr_c \pi}).
\end{equation}
There is no hierarchy problem since all the parameters can be 
of the order of the Planck scale.  
In
Ref. \cite{RS1} (referred to as RS1 scenario), the Standard Model
particles are confined on the boundary at $\phi=\pi$. To explain the
electroweak scale, $r_c$ is chosen as $k \exp(-kr_c \pi)\sim $ TeV. In
Ref. \cite{LR} (referred to as LR scenario), there appear three
3-branes, one is the Planck brane at $\phi=0$, another one with
negative tension is taken away to infinity $r_c \rightarrow
\infty$. And the Standard Model particles are on the other brane at
finite $\phi=\phi_0$. There will be  a physical scale $\Lambda= 
k \exp(- k r_c \phi_0)\sim$ TeV, which leads
$k r_c \phi_0 \sim 35$.  This is an interesting idea since the Standard
Model physics below TeV scale will not depend on the total size of the fifth
dimension, so the one boundary of the extra dimension can be located
at an arbitrarily distant place.

In this letter, we would like to examine this idea in more detail. The
situation we shall argue is similar to the LR model, but a finite
$r_c$ is kept. We will show that the size of the extra dimension $r_c$ 
plays an important role in physics of the KK modes of the graviton.

Following the conventions of Ref. \cite{DHR}, the interaction  
between the Standard Model particles  and the bulk graviton  is 
written
\begin{equation}
{\cal L}=-\frac{1}{M^{3/2}} T^{\mu\nu} h_{\mu\nu}(x,\phi_0),
\end{equation}
where $T^{\mu \nu}$ is an energy-momentum tensor of the Standard Model and 
the metric fluctuation $h_{\mu \nu}$ can be expanded as follows:
\begin{equation}
h_{\mu\nu}(x,\phi)=\sum^{\infty}_{n=0} h_{\mu\nu}^{(n)}(x)\frac{\chi^{(n)}
(\phi)}{\sqrt{r_c}}.
\end{equation}
Here the mode function is normalized as $\int d \phi \exp (kr_c \phi)
\chi^{(m)}(\phi) \chi^{(n)} (\phi)=\delta_{mn}$. 
Solving an equation of motion for each mode function, we find that it is 
expressed in Bessel functions
\begin{equation}
\chi^{(n)}(\phi)=\frac{e^{2 k r_c\phi}}{N_n}\left[J_2(z_n) + \alpha_n
Y_2(z_n)\right],
\end{equation}
where $z_n(\phi)= m_n e^{kr_c\phi} /k$. The boundary condition at $\phi=0$ 
implies that the coefficient $\alpha_n$ is negligible. On the other hand,
the boundary condition at $\phi=\pi$ yields $J_1(z_n(\pi))\simeq 0$, which
determines the masses of the KK modes. For $z_n(\pi) \gg 1$, this becomes 
simply  $z_n(\pi)-3\pi/4=\pi \times$(integer$+\frac{1}{2}$) and thus the mass gap is 
\begin{equation}
\Delta m=m_{n+1}-m_n=\pi k e^{-k r_c \pi}.
\end{equation}
Finally one finds the normalization 
\begin{equation}
N_n \simeq \frac{e^{kr_c\pi}}{\sqrt{kr_c}} |J_2(z_n(\pi))|.
\end{equation}
In the asymptotic limit $z_n(\pi)\gg 1$, the above reads
\begin{equation}
N_n \simeq \frac{e^{kr_c\pi}}{\sqrt{kr_c}} \sqrt{\frac{2}{\pi z_n(\pi)}}.
\end{equation}
With the approximation for small $z_n(\phi_0)$, $J_2(z_n(\phi_0))\simeq
z_n(\phi_0)^2/8$,
the coupling of each KK mode with mass $m_n$ to the Standard Model
particles  is obtained
\begin{equation}
\frac{\chi^{(n)}(\phi_0)}{M^{3/2} \sqrt{r_c}} \simeq
\frac{(\Delta m)^{1/2}}{8\sqrt{2}}\left(\frac{k}{M}\right)^{3/2} 
\frac{m_n^{5/2}}{\Lambda^4}.
\end{equation}


A crucial point here is that the decay width of a graviton KK mode is 
proportional
to the mass gap, characterized by the location of the negative-tension brane. 
In fact, the total decay width for each KK mode is estimated to be \cite{han2}
\begin{equation}
\Gamma(m_n) \simeq 3\times 10^{-5} N \left(\frac{k}{M}\right)^3 
\frac{m_n^8}{\Lambda^8}
\Delta m,
\end{equation}
with a coefficient $N$ parameterizing a number of channels with appropriate 
weight, which in the following we take $N \sim 10$. 
Here we would like to emphasize the physical importance of the size of
the extra dimension $r_c$ in cosmology and in phenomenology. When $r_c$ is 
infinitely large as in the LR scenario, $\Delta m$ is 
infinitesimal and hence the lifetime becomes infinity. Thus in this case the
KK mode is absolutely stable.  As the $r_c$ becomes small, the lifetime
becomes cosmologically relevant. For a much smaller $r_c$, the lifetime of
the KK mode is so small that it decays even inside a detector in a collider
experiment. 

In the early universe, the 
KK modes will be produced through thermal interactions.
When the reheat temperature (or the normalcy temperature defined in 
Ref. \cite{ADD1}) is $T$, the yield $\Delta Y(T)$ for a KK mode with mass $m_n$
produced from various
processes is evaluated as, 
\begin{equation}
\Delta Y(T)\simeq \frac{\Gamma(m_n)}{H(T)} Y_{EQ}\simeq
 10^{-6}\tilde{N}  
\frac{M_{pl}}{T^2}\left(\frac{k}{M}\right)^3
\frac{m_n^8}{\Lambda^8} \Delta m,
\end{equation}
where $Y_{EQ}\simeq0.3/g_{*s}$ is a yield in equilibrium and
we  can keep the order one constant $\tilde{N}$  for reasonable
temperature ranges.

When $r_c\rightarrow \infty$, {\it i.e.} each KK mode is absolutely stable,
we can calculate the (mass density)/(entropy density) of the sum over KK modes
\begin{equation}
\frac{\rho}{s}\simeq\int m\times  dY
\simeq  10^{-7}\tilde{N} M_{pl} \left(\frac{k}{M_{pl}}\right)^2
\left(\frac{T}{\Lambda}\right)^8.
\end{equation}
To avoid the over-closure of the universe, $\rho/s$ at the present time 
should be
less than $3\times 10^{-6} h^2$ MeV,
which gives a bound on the normalcy temperature of the universe
\begin{equation}
T< 3\cdot 10^{-3} \Lambda \sim \mbox{ GeV}.
\end{equation}

On the other hand, as $r_c$ becomes small, the decay of the graviton KK mode
becomes of cosmological relevance. Since the decay products generally
contain radiation, the decay is potentially cosmological embarrassment, from
which one can get limits on the abundances of the graviton KK modes. Here
we will not go into detail any more; rather we would like to point out another
extreme case where all KK modes decay much before the big-bang
nucleosynthesis started.  This is the case when the life time of the lightest
KK mode with mass $\Delta m$ is much shorter than 1 sec, namely
\begin{equation}
\Gamma^{-1}\simeq 10^4 
\left(\frac{M}{k}\right)^3 \frac{\Lambda^8}{\Delta m^9}
\ll 1 \mbox{ sec},
\end{equation}
or
\begin{equation}
\Delta m \gg 3 \left(\frac{M}{k}\right)^{1/3} \left(\frac{\Lambda}{\mbox{TeV}}\right)^{8/9}
\mbox{ GeV} .
\end{equation}
Note that $\Delta m \sim 1$ GeV corresponds to $k \pi r_c \sim 40$,
which is slightly larger than $k r_c\phi_0\sim 35$.
If $\phi_0\simeq \pi$, the scenario of Ref.\cite{RS1} shares the same property with this case since
all KK modes have masses of the order $\Lambda$ and higher, and thus are
very short lived.

Let us discuss collider signatures in these models. Here we
examine some typical cases. In the RS1 scenario, the mass gaps between
the KK modes are of  order $\Lambda$ and so the cleanest signal will be
given by resonances of the KK modes \cite{DHR}. As the size of the extra
dimension gets larger, the mass gap becomes smaller, and then it will
be difficult to distinguish each resonance peak; rather the signal
will be of contact interaction type. Indeed the sum over the virtual KK modes
gives the four Fermi interaction proportional to
\begin{equation}
\int dm \frac{m^5}{\Lambda^8} \frac{1}{s-m^2}
\sim \frac{1}{\Lambda^2},
\end{equation}
where we have introduced a cut-off around $\Lambda$. We expect that it will
be provided by taking into account fluctuations of the Standard-Model brane
\cite{Bando}. 
In either case the spin 2 nature
of the graviton KK modes is seen from the angular distribution of the 
final states. In the LR limit where these KK modes are stable, another type
of signal will be expected. Namely the KK modes are emitted from
the gauge bosons (photon in a lepton collider and gluon in a hadron collider),
which escape from detection. The signature of the LR limit is similar
to the case of the six extra dimensions in Ref. \cite{ADD1}.  When the mass gap
is large, the decay length of the KK mode will be finite. For a certain 
range of the parameters we may be able to see a displaced vertex from the 
collision point, which consists of the KK graviton decaying to charged 
particles. It is worth mentioning that  if the mass gap is around $10^{-3}$ 
eV, which could be identified with the energy scale of  seemingly 
non-vanishing cosmological constant, the decay length of the KK mode with
 mass $\sim \Lambda$ will be of order 1 m 
and thus this case would provide a spectacular signal of the displaced vertex.
For a much smaller mass gap like in the case of $\Delta m \sim$ 1 GeV as we 
discussed above, the decay length is too short to be detected, and thus
this type of signals will be diminished.
 

To summarize, we have investigated the coupling of the graviton KK mode
to Standard Model particles in the warped extra dimension where the exponential
dependence of the metric on the location in the extra dimension gives rise
to the electroweak scale.  
We pointed out the length of the fifth dimension is of
physical importance as it crucially affects to the decay widths of KK
modes of the graviton. For the infinitely large extra dimension as in
the LR model, the graviton KK modes are absolutely stable, whereas 
they become short lived as the length of the extra dimension becomes smaller.
We argued the implications of this fact to cosmological
constraints on the KK graviton physics and also to searches of the KK
modes at collider experiments and showed that both of them strongly depend
on the length of the fifth dimension. More details on the collider signatures
and on the cosmological constraints will be discussed elsewhere.

\

We would like to acknowledge J. Hisano, K.-I. Izawa, T. Kugo, H. Nakano,
Y. Nomura and A. Pomarol for delightful conversations during the
summer institute 99 at Yamanashi, Japan.  
This work was supported in part by the Grant-in-Aid for Scientific 
Research from the Ministry of Education, Science, Sports, 
and Culture of Japan, 
on Priority Area 707 ``Supersymmetry and Unified Theory of Elementary
Particles", and by the Grant-in-Aid No.11640246 and  No.98270.
SC thanks the Japan Society for the Promotion of Science for financial support.

\end{document}